%% file: PlectonemePaperShort.tex
\pdfoutput=1
\documentclass[prl,superscriptaddress,twocolumn,floatfix]{revtex4}

\usepackage{amsmath}
\usepackage{graphicx}

\newcommand{\little}{2.in}
\newcommand{\figref}[1]{Fig.~\ref{#1}}

\renewcommand{\eqref}[1]{Eq.~(\ref{#1})}
\newcommand{\eqrefTwo}[2]{Eqs.~(\ref{#1}) and (\ref{#2})}
\newcommand{\degree}{$^\circ$}

\newcommand{\F}{\mathcal{F}}
\newcommand{\G}{\mathcal{G}}
\newcommand{\str}{\mathrm{SS}}
\newcommand{\coex}{\mathrm{CS}}
\newcommand{\Wr}{\mathrm{Wr}}
\newcommand{\Loop}{\mathrm{loop}}
\newcommand{\Feff}{F_{\mathrm{eff}}}

\begin{document}

\title{Discontinuities at the DNA supercoiling transition}
\date{\today}
\author{Bryan C.~Daniels}
\author{Scott Forth}
\author{Maxim Y.~Sheinin}
\affiliation{Department of Physics, 
Laboratory of Atomic and Solid State Physics, Cornell University,
Ithaca, NY 14853}
\author{Michelle D.~Wang}
\affiliation{Department of Physics, 
Laboratory of Atomic and Solid State Physics, Cornell University,
Ithaca, NY 14853}
\affiliation{Howard Hughes Medical Institute, Cornell University,
Ithaca, NY 14853}
\author{James P.~Sethna}
\affiliation{Department of Physics, 
Laboratory of Atomic and Solid State Physics, Cornell University,
Ithaca, NY 14853}
\begin{abstract}
While slowly turning the ends of a single molecule of DNA at constant
applied force, a discontinuity
was recently observed at the supercoiling transition, when a small
plectoneme is suddenly formed.
This can be understood as an abrupt transition into a state in which
stretched and plectonemic DNA coexist.  We argue that there
should be discontinuities in both the extension and the torque at the
transition, and provide experimental evidence for both.  To predict the 
sizes of these discontinuities and how they change with the
overall length of DNA, we 
organize a phenomenological theory for the coexisting plectonemic state
in terms of four parameters.
We also test supercoiling theories, including our own elastic rod simulation,
finding discrepancies with experiment that can be understood in terms 
of the four coexisting state parameters.  
\end{abstract}
\maketitle

\begin{figure}
\centering
\includegraphics[width=\linewidth]{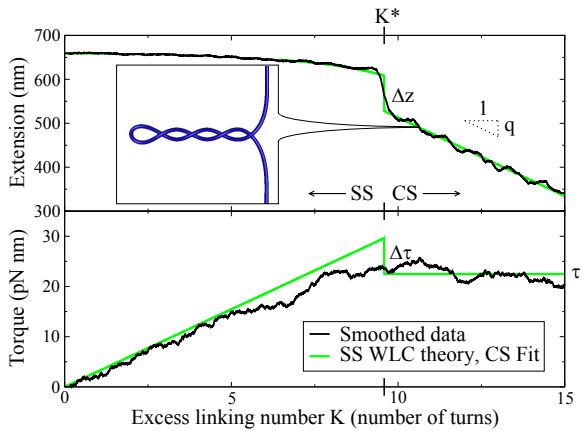}
\caption{\label{fig:ExtensionAndTorqueVsK}%
Extension and torque as a function of linking number $K$, for $L=2.2$ kbp
at $F=2$ pN.  
Black lines show data from Ref.~\cite{ForDeuShe08}, smoothed using 
a ``boxcar'' average of nearby points.
The green lines show worm-like chain (WLC) predictions below 
the transition [in the unsupercoiled ``stretched'' state (SS)], and 
fits to the data after the transition [in the 
``coexisting'' state (CS)], linear for
the extension and constant for the torque.  
The size of the torque jump, not visible in the smoothed
data, is implied by the coexisting torque $\tau$, the CS fit, and the transition 
linking number $K^*$ in the extension data.  
Inset: Simulated DNA showing the CS of a plectoneme and straight DNA,
ignoring thermal fluctuations.
The ends are held with fixed orientation and pulled with a constant force
$F$, here 2~pN.
}
\end{figure}

\begin{figure}
\centering
\includegraphics[width=\linewidth]{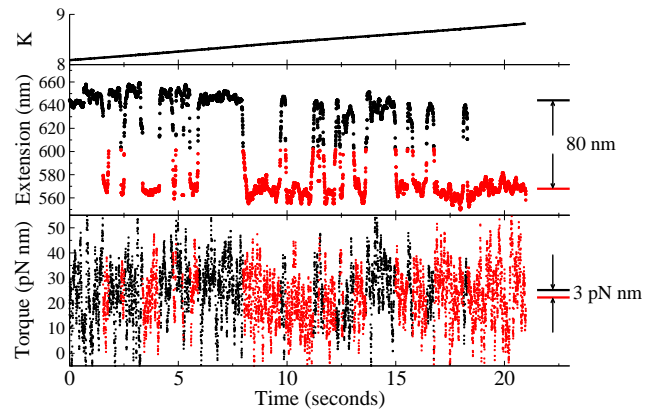}
\caption{\label{fig:Hopping}%
Directly measuring the torque jump by observing thermal hopping, for the
same conditions as \figref{fig:ExtensionAndTorqueVsK}.  As linking
number $K$ is slowly increased near $K^*$, thermal fluctuations induce 
hopping between
states with (CS) and without (SS) a plectoneme.  
Averaging over these two states gives a direct way of measuring 
the torque jump: analogously to a lock-in amplifier, we
set a threshold in the extension signal to
separately average the SS (black) and CS (red) data near the 
transition.  
Using multiple traces, we find an average torque jump
of $\Delta \tau = 2.9 \pm 0.7$ pN~nm for $L=2.2$ kbp at $F=2$ pN.  
Additionally, this
value of $\Delta \tau$ implies (see text) that the transition should
happen over a range of linking number $K$ (top) of about 0.9 turns, 
as it does.
}
\end{figure}

A DNA molecule, when overtwisted, can form a 
{\em plectoneme}~\cite{StrAllBen96,CruKosSei07} 
(inset of \figref{fig:ExtensionAndTorqueVsK}), 
a twisted supercoil structure familiar from phone cords and water
hoses, which 
stores added turns (linking number) as `writhe.'  The plectoneme
is not formed when the twisted DNA goes unstable (as in water
hoses~\cite{HeiNeuGos03}),  
but in equilibrium when the free energies cross --- this was vividly 
illustrated by a recent experiment~\cite{ForDeuShe08} 
(\figref{fig:Hopping}), which showed repeated transitions between
the straight ``stretched state'' (SS, described by the worm-like chain
model~\cite{MarSig95b}), and a coexisting state (CS) of stretched DNA and 
plectoneme~\cite{Mar07}.
This transition, in addition to being both appealing and biologically
important, provides an unusual opportunity for testing continuum theories
of coexisting states.
Can we use the well-established continuum theories of DNA elasticity to
explain the newly discovered~\cite{ForDeuShe08} jumps in behavior at
the transition?

The recent experiment measures the extension (end-to-end distance) and torque
of a single molecule of DNA held at constant force as it is slowly twisted
\cite{ForDeuShe08}.
A straightforward numerical implementation of the elastic rod 
model~\cite{FaiRudOst96,MarSig95a,Neu04} for DNA in 
these conditions
(with fluctuations incorporated via entropic repulsion~\cite{MarSig95a}) leads 
to two quantitative predictions that are at variance with the experiment.
First, the experiment showed a jump $\Delta z$ in the extension
as the plectoneme formed (\figref{fig:ExtensionAndTorqueVsK}) that 
appeared unchanged for each applied force as the overall DNA length 
was varied from 2.2~kbp to 4.2~kbp, whereas the
simulation showed a significant increase in $\Delta z$ at the 
longer DNA length.
Second, no discontinuity was observed in the (directly measured) filtered
torque data (\figref{fig:ExtensionAndTorqueVsK}), yet the simulation
predicted a small jump.

Simulation is not understanding. Here we analyze the system theoretically,
focusing on the physical causes of the behavior at the transition.
We use as our framework Marko's two-phase coexistence model~\cite{Mar07,Mar08},
which we generalize to
incorporate extra terms that represent the interfacial energy between
the plectoneme and straight regions of the DNA. We show that any model
of the supercoiling transition in this parameter regime can be summarized
by four force-dependent parameters.  After extracting these parameters directly
from the experiments, we use them to predict the torque
jump (which we then measure) and to explain why the extension jump appears 
length independent. 
Finally, we use our formulation to test various models 
of plectonemes, finding discrepancies mainly at small applied force.

The transition occurs at the critical linking number $K^*$ 
when the two states have the same free energy $\F$,
where $\F$ is defined by the ensemble with constant 
applied force and linking number. We therefore need models for
the free energy $\F$ and extension $z$ of the SS and CS.

The properties of stretched, unsupercoiled DNA are 
well-established.  At small enough forces and torques that
avoid both melting and supercoiling,
DNA acts as a torsional spring with twist elastic constant $C$
\cite{Mar07}\footnote{
	As described in Ref.~\cite{MorNel97} (see also supplemental material), $C$ is
	renormalized to a smaller value by bending fluctuations.  We
	use $C$ calculated from the torque measured in the experiment, which
	gives its renormalized value.
}:
$\F_\str(K,L) = \frac{C}{2} \left( 2\pi \frac{K}{L} \right)^2 L - \Feff L$,
where $K$ is the added linking number, $L$ is the overall (basepair) length
of DNA, the effective force $\Feff  = F - kT\sqrt{F/B}$ \cite{Mar07} (see
supplemental material), 
$F$ is the force applied to the ends of the DNA,
$B = 43 \pm 3$~nm$\times kT$ is the DNA's bending elastic constant,
$C$ = $89 \pm 3$~nm$\times kT$, and the thermal
energy $kT=$~4.09~pN~nm for this experiment (at 23.5\degree C).  
Differentiating with respect
to $K$ gives the torque:
$\tau_\str = \frac{1}{2\pi} \frac{d \F_\str}{dK} = 2\pi C \frac{K}{L}$.
The extension of unsupercoiled DNA is shortened by thermal fluctuations,
and in the relevant force regime is approximately given by 
$z_\str = \xi(\tau_\str) L$,
where \cite{MorNel98} 
\begin{equation}
\label{xi}
\xi(\tau) = 1 - \frac{1}{2} \left[ \frac{B F}{(k T)^2} 
 				-\left(\frac{\tau}{2 k T}\right)^2
 				-\frac{1}{32} \right]^{-1/2}.
\end{equation}




Since supercoiling theories must include contact forces,
they are less amenable to traditional theoretical methods.
Even so, many theories have been successful
in predicting properties of the CS;
such methods have included detailed Monte
Carlo simulations \cite{VolMar97}, descriptions of
the plectoneme as a simple helix \cite{MarSig95a,Neu04,ClaAudNeu08}, 
and a more phenomenological approach \cite{Mar07}.
However, none of these theories has yet been used to predict 
discontinuities at the SS--CS transition.
Here we 
connect the free energy and extension predictions from any
given model to the corresponding predictions for discontinuities
at the transition.

We will use the framework of two-phase coexistence 
adopted by Marko \cite{Mar07,Mar08} to describe the
CS as consisting of two phases, each 
with constant free energy and extension per unit length
of DNA \footnote{
	The language of phase coexistence is approximate in that 
	the finite barrier to
	nucleation in one-dimensional systems precludes a true
	(sharp) phase transition.
}.  Since phase coexistence leads to a linear dependence on $K$ of the
fraction of plectonemic DNA (keeping the torque fixed), in this model both
$\F_\coex$ and $z_\coex$ are linear functions of added linking
number $K$ and length $L$ (just as the free energy of an ice-water mixture
is linear in the total energy, and the temperature remains fixed, as the
ice melts).
This linearity, along with the known properties of the SS, 
allows us to write $\F_\coex$ and $z_\coex$ as
(see supplemental material)
\begin{align}
\label{F_coex}
\F_{\coex}(K,L) &= \F_0 
	+ 2\pi \tau K - \left( \frac{\tau^2}{2C} + \Feff  \right) L ; \\
\label{z_coex}
z_{\coex}(K,L)  &= -z_0 
    - q K 
    + \left( \xi(\tau) + \frac{\tau}{2\pi C}q \right)L,
\end{align}
where $q$ is the slope of extension versus linking number and $\tau$
is the CS torque.
That is, $\F_\coex$ and $z_\coex$ are specified by four force-dependent 
values: their slopes with respect to $K$
($\tau$ and $q$), which describe how the plectonemic phase 
coexists with the stretched phase; and $K=L=0$ offsets 
($\F_0$ and $z_0$), 
which describe the extra free energy and 
extension necessary to 
form the interface between the phases --- the end loop and tails of 
the plectoneme.

The experimental observables can then be written in terms of these
four values.   
Easiest are $\tau$ and $q$,
which are directly measured.
Next, the linking number $K^*$ at the transition is found by
equating the CS free energy with that of the SS:
$\F_{\coex}(K^*,L) = \F_\str(K^*,L)$ implies
\begin{equation}
\label{Kstar}
K^* = \frac{L}{2\pi C}(\tau + \Delta \tau),
\mathrm{with}~\Delta \tau = \sqrt{ \frac{2C}{L} \F_0 },
\end{equation}
where $\Delta \tau$ is the jump in the torque at the transition.
Lastly, inserting $K^*$ from \eqref{Kstar} into \eqref{z_coex}, we  
find the change in extension at the transition:
\begin{align}
\Delta z &= 
	z_0 + q \sqrt{\frac{L\F_0}{2\pi^2 C} }
	    - L \Big( \xi(\tau) - \xi( \tau + \sqrt{ 2C\F_0/L }) \Big).
	 \label{Deltaz}
\end{align}

\begin{figure}
\centering
\includegraphics[width=\linewidth]{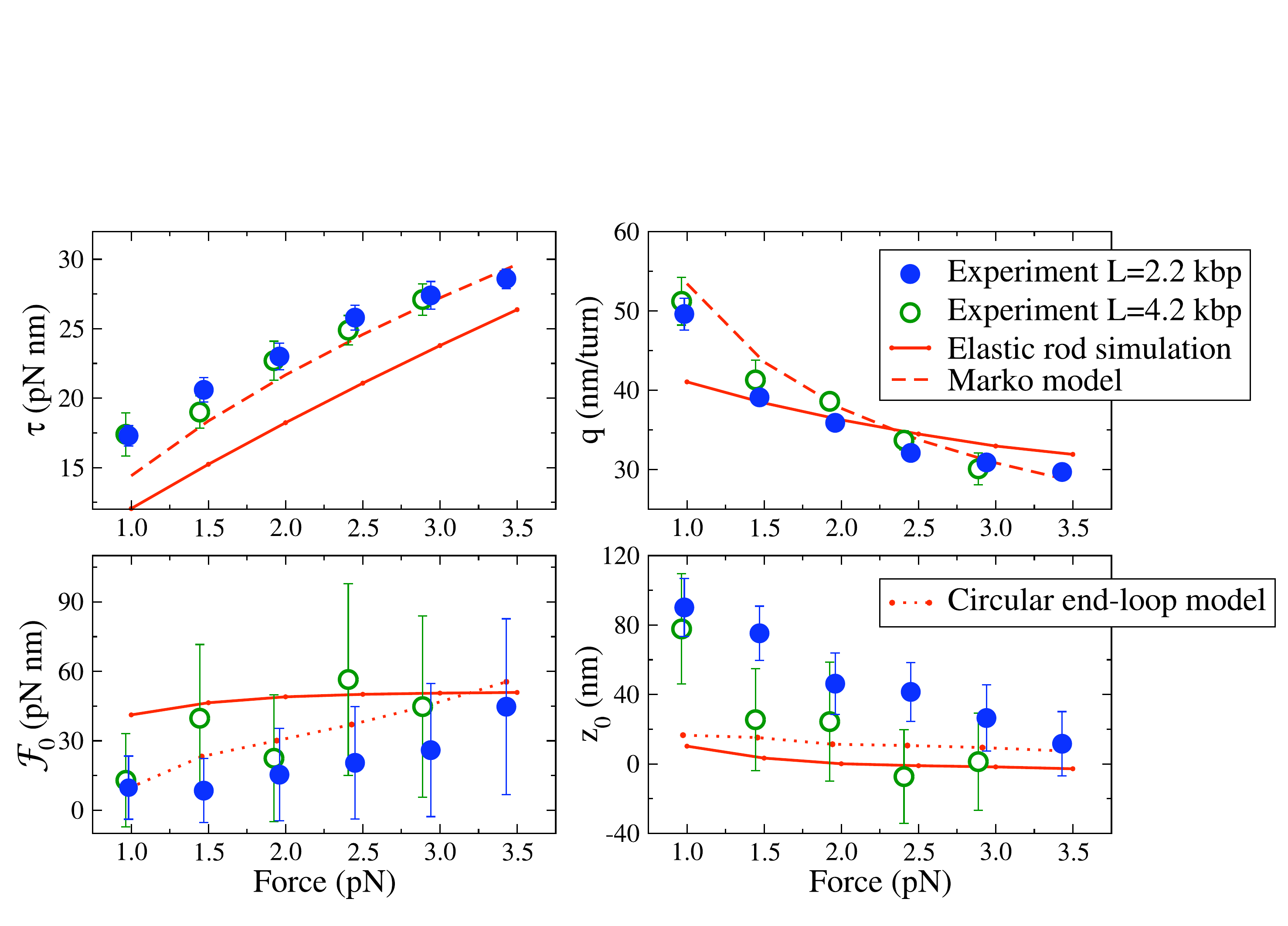}
\caption{\label{fig:FourPlots}%
The four parameters describing the CS (coexisting torque $\tau$, 
extension versus linking number slope $q$, and the extra free
energy $\F_0$ and extension $z_0$ necessary to
form the end loop and tails of the plectoneme), as a function
of applied force.  The circles show values calculated from 
experimental data taken at two different overall DNA lengths $L$.
Model predictions for our simulation
\cite{endnote20}
and Marko's model \cite{Mar07} are 
shown as solid and dashed lines, respectively (using $S=0$ for $\F_0$
predictions).  The circular end-loop model uses
average $\tau$ and $q$ values from the experiment to predict
$\F_0$ and $z_0$, shown as dotted lines. 
}
\end{figure}

\begin{figure}
\centering
\includegraphics[width=\linewidth]{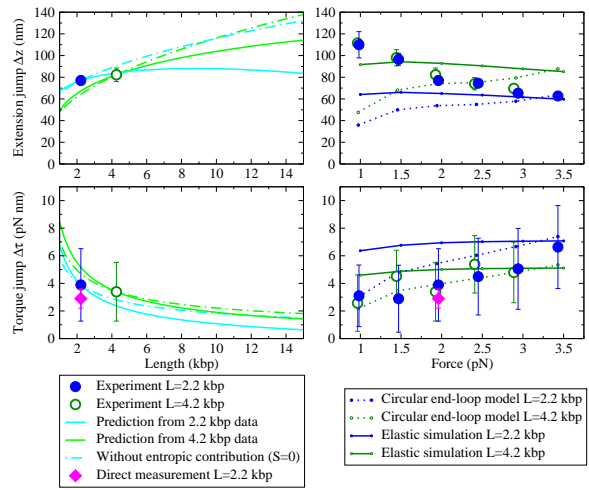}
\caption{\label{fig:Discontinuities}%
(Left) Predicted length-dependence of the extension and torque jumps at
$F=2$ pN.
Using the CS parameters extracted from the experiment at two different lengths,
Eqs.~(\ref{Kstar}) and (\ref{Deltaz}) predict the $L$-dependence of 
$\Delta z$ and $\Delta \tau$.  
The circles show experimentally-measured
values [with the torque jump here calculated from $K^*$
using \eqref{Kstar}].  Without entropic corrections to $\F_0$ 
($S=0$; dot-dashed lines) $\Delta z$ depends noticeably on $L$, but
including an initial estimate of $S$ (solid lines) shows that entropic 
effects can significantly reduce this length-dependence.
(Right) Force-dependence of the extension and torque jumps, and predictions
from two models.  
Disagreements with experimental data can be understood in terms of
the four CS parameters in \figref{fig:FourPlots}.  
Also plotted as a diamond is $\Delta \tau$ measured using the
direct method depicted in \figref{fig:Hopping}. 
}
\end{figure}

%


To additionally include entropic effects, we can write $\F_0 = \mu - TS$, 
where $\mu$ is
the energy cost for the end-loop and tails, and $S$ is the entropy coming
from fluctuations in the location, length, and linking number of the 
plectoneme.  Using an initial calculation of $S$ that includes these
effects (in preparation; see supplemental material), we 
find that $S$ varies logarithmically with $L$, and that setting $S=0$ is
a good approximation except when $L$ changes by large factors.

Given experimental data ($\tau$, $q$, $K^*$, and $\Delta z$), 
we can solve for the four
CS parameters.  The results from Ref.~\cite{ForDeuShe08}
are shown as circles in \figref{fig:FourPlots} for the two overall DNA lengths
tested.  If we assume that the DNA is homogeneous, we expect the results 
to be independent of $L$ (except for a logarithmic entropic correction to 
$\F_0$ that would reduce it at the longer $L$ by about
$kT \log 2 \approx 5$ pN~nm; see supplemental material).  
We do expect $\F_0$ and $z_0$ to be sensitive to the local 
properties of the DNA in the end-loop of the plectoneme, so we 
suspect that the difference in $z_0$ between the two measured 
lengths could be due to sequence dependence.
With this data, we can 
also predict the length-dependence of the discontinuities, as shown
in \figref{fig:Discontinuities} (left).  
Here we included entropic corrections to $\F_0$ (see supplemental
material), and we find that  
entropic effects significantly decrease the length-dependence of 
the extension jump.

Note that here we are solving for the experimental size of the
torque jump using the observed $K^*$ and $\tau$ in \eqref{Kstar}.
We also find direct evidence of $\Delta \tau$ in the data by 
averaging over the torque separately in the SS and CS near the transition 
(\figref{fig:Hopping}).  With data taken at $F=2$ pN and $L=2.2$ kbp, 
we find $\Delta \tau = 2.9 \pm 0.7$ pN~nm, in good 
agreement with the prediction from $K^*$ 
($3.9 \pm 2.6$ pN nm; see \figref{fig:Discontinuities}).
We can also predict the width of the range of linking numbers around
$K^*$ in which hopping between the two states is likely 
(where $|\Delta \F|<kT$):
expanding to first order in $K-K^*$ gives a 
width of $2 kT / (\pi \Delta \tau$).  This predicts a transition region
width of about 0.9 turns for the conditions in
\figref{fig:Hopping}, agreeing well with the data.


We can now use various plectoneme models to calculate the four CS parameters, 
which in turn give predictions for the experimental observables.
The results are shown as lines in \figref{fig:FourPlots} and 
\figref{fig:Discontinuities} (right).  As we expect entropic corrections to 
be small (changing $\F_0$ by at most about 5 pN nm), we set $S=0$ 
for these comparisons.

First, we test Marko's phase coexistence model \cite{Mar07}.  
The plectoneme is modeled as a phase with
zero extension and an effective twist stiffness $P < C$.  Shown as dashed lines
in \figref{fig:FourPlots}, the Marko model predicts the coexisting 
torque and extension slope well, with $P$ as the only fit parameter
(we use $P=26$~nm).  
However, 
the Marko model (and any model that includes only terms
in the free energy proportional to $L$) produces $\F_0=0$ and $z_0=0$.

In order to have a discontinuous transition, we must include the 
effects of the end loop and tails of the plectoneme.  
The simplest model assumes that the coexistence of stretched and
plectonemic DNA requires one additional circular loop of DNA.  Minimizing
the total free energy for this circular end-loop model gives
\begin{align}
\F_0 &= 2\pi \sqrt{2B\Feff } - 2\pi \tau \Wr_\Loop;		\label{circularz0} \\
z_0 &= 2\pi \xi(\tau) \sqrt{B/(2\Feff )} - q\Wr_\Loop,	\label{circularF0}
\end{align}
where $\Wr_\Loop$ is the writhe taken up by the loop. For a
perfect circle, $\Wr_\Loop=1$, and $\Wr_\Loop<1$ for a loop with two ends
not at the same location.  We chose $\Wr_\Loop=0.8$ as a reasonable best fit 
to the data.  Using the experimentally measured $\tau$ and $q$, 
the predictions are shown as solid
lines in \figref{fig:FourPlots} and \figref{fig:Discontinuities}; $\F_0$ 
is fit fairly well, but $z_0$ is underestimated, especially at small applied
forces. 

In an attempt to more accurately model the shape of the plectoneme,
we use an explicit 
simulation of an elastic rod, with elastic constants
set to the known values for DNA.  We must also include repulsion between
nearby segments to keep the rod from passing through itself.  Physically,
this repulsion has two causes: screened Coulomb interaction of the charged 
strands and the loss of entropy due to limited fluctuations in the
plectoneme.  We use the repulsion free energy derived for the helical
part of a plectoneme
in Ref.~\cite{MarSig95a}, modified to a pairwise potential form (see
supplemental material).
We find that the simulation does form 
plectonemes (inset of \figref{fig:ExtensionAndTorqueVsK}), and we 
can extract
the four CS parameters, shown as solid lines in
\figref{fig:FourPlots}
\footnote{ 
	We have also explored increasing the
	entropic repulsion by a constant factor of up to 3.  
	Though this does bring the torques closer to the experiment, 
	the only other significant change 
    is a decrease in $\F_0$ (data not shown) --- specifically, 
	this does not change the discussed discrepancies between
	the simulation and experiment. 
}\label{footnote:RodSimulationChoice}.  Since $\F_0$ and $z_0$ are nonzero, we
find discontinuities in the extension and torque at the transition;
their magnitudes are
plotted in \figref{fig:Discontinuities}.

Both the circular loop model and the simulation produce torque and 
extension jumps of the correct 
magnitude, but in both cases $\Delta z$ has an incorrect 
dependence on force and too much dependence on length. 
Our approach provides intuition about the causes
of the discrepancies by singling out the four values 
(connected to different physical effects) that combine to produce the 
observed behavior. 
Specifically, we can better understand why the models' predictions are
length-dependent: as displayed in
\figref{fig:Discontinuities} (top left), 
the negligible length-dependence 
observed in experiment is caused by a subtle cancellation of a 
positive length-dependence [smaller than either model, 
and described by \eqref{Deltaz}] combined with a negative contribution 
coming from entropic effects.  One would expect, then, that
any plectoneme model (even one that explicitly 
includes entropic fluctuations) might easily miss this cancellation.  
In general, without this intuition, 
it is difficult to know where to start in improving the DNA models.


The largest uniform discrepancy
happens at small applied forces, where both models underestimate
$z_0$ \footnote{
	Though the 4.2 kbp $z_0$ data alone would be arguably consistent with
	the model predictions, the 2.2 kbp data highlights the discrepancy at
	small applied forces. 
}, leading to an underestimate 
of $\Delta z$.  We have examined various 
effects that could alter $z_0$,
but none have caused better agreement (see also supplemental material).
Adding to the circular end-loop model softening or kinking \cite{DuKotVol08} 
at the plectoneme tip, or entropic 
terms from DNA cyclization theories \cite{ShiYam84,Odi96},  
uniformly {\em decreases} $z_0$.  
Increasing $B$ in \eqref{circularz0} by a factor of four (perhaps
due to sequence dependence) does raise $z_0$ into the correct range,
but it also raises $\F_0$ from \eqref{circularF0} to values well outside the 
experimental ranges.   
Finally, $z_0$ would be increased if multiple 
plectonemes form at the transition, 
but we find that
the measured values of $\F_0$ are too large to allow for more than one 
plectoneme in this experiment.




Support is acknowledged from NSF Grants DMR-0705167 and MCB-0820293,
NIH Grant GM059849, and the Cornell Nanobiotechnology Center.



\bibliographystyle{apsrev}
\bibliography{PlectonemePaper}

\include{PlectonemeSupplementalInfo}

\end{document}

%% file: PlectonemeSupplementalInfo.tex
\renewcommand{\little}{4.in}
\newcommand{\Fex}{\F_{\mathrm{extra}}}
\newcommand{\Ceff}{C_{\mathrm{eff}}}



\renewcommand{\thepage}{S\arabic{page}}
\setcounter{page}{1}
\renewcommand{\theequation}{S\arabic{equation}}
\setcounter{equation}{0}
\renewcommand{\thefigure}{S\arabic{figure}}
\setcounter{figure}{0}

\begin{widetext}

\begin{center}
\Large{Supplemental Material} \\ ~ \\ 
    \large{\emph{Discontinuities at the DNA supercoiling transition}}

Bryan C.~Daniels, Scott Forth, Maxim Y.~Sheinin, 
		Michelle D.~Wang, James P.~Sethna \\ ~ \\
\end{center}

\subsection{Behavior of extended DNA with fluctuations}
\label{appendix_fluctuatons}

The behavior of extended DNA is appreciably affected by thermal
fluctuations.  For the applied forces in the range considered in this experiment, 
we can use the following
fixed-torque free energy:
\begin{equation}
\label{eq_g}
\frac{\G(\tau)}{L} = - F - \frac{\tau^2}{2 C_{bare}} 
	+ \frac{kT}{B}\sqrt{BF - \frac{\tau^2}{4}},
\end{equation}
where the last term is the lowest-order correction
due to fluctuations \cite{MorNel98}. 

The fluctuations decrease the extension:
\begin{equation}
-\frac{\partial \G}{\partial F} = 
  L \left[ 1 - \frac{kT}{2} \left( BF - \frac{\tau^2}{4} \right)^{-1/2} \right].
\end{equation}
(The $-1/32$ in Eq.~(1) comes from an approximation 
to a higher-order correction \cite{MorNel98}.)

Expanding the last term of \eqref{eq_g} to match the form of 
a ``zero-temperature'' chain, we can instead write
\begin{equation}
\label{eq_g_eff}
\frac{\G(\tau)}{L} = - \Feff  - \frac{\tau^2}{2 \Ceff },
\end{equation}
where the effective force and twist elastic constant are
given by
\begin{align}
\Feff  &= F - kT\sqrt{\frac{F}{B}} \\
\Ceff  &= C_{bare} \left( 1 + kT\frac{C_{bare}}{4B\sqrt{BF}} \right)^{-1}.
\end{align}
Note that $\Ceff $ is a function of force: there is less ``softening''
at higher forces.  In the experiments of Forth \textit{et al.}, the renormalized
$\Ceff $ was measured directly via the torque.  However, the
range of applied forces was small enough that $\Ceff $ did not 
change appreciably, and a single value of $C=(89~\mathrm{nm})kT$ was quoted.
Here, we also use the same renormalized but 
force-independent value for $C$.

Changing \eqref{eq_g_eff} to a fixed-linking-number expression via a Legendre 
transformation, we arrive at our expression for the straight
state free energy (also found in Ref.~\cite{Mar07}):
\begin{equation}
\F_s(K,L) = \frac{C}{2} \left( 2\pi \frac{K}{L} \right)^2 L - \Feff L.
\end{equation}

\subsection{Derivation of linear expressions for $\F_\coex$ and $z_\coex$}



We first write down the linear scaling of the free energy
and extension with linking number.
For any $\delta K$
that does not take the system out of the CS,
\begin{align}
\label{FKscaling}
\F_{\coex}(K+\delta K,L)&= \F_{\coex}(K,L) + 2\pi \tau \delta K ;\\
\label{zKscaling}
z_\coex(K+\delta K,L)	&= z_\coex(K,L) - q \delta K,
\end{align}
where $q$ is the slope of extension versus linking number and $\tau$
is the CS torque.
Next, to find the scaling with increasing $L$, we imagine adding 
a piece of stretched DNA of length $\delta L$ at the coexisting 
torque (keeping the system in a stable CS).  
This also adds an amount of linking number 
that scales with $\delta L$,
$\delta K [\delta L] = \tau \delta L / (2\pi C) $,
which we will have to unwind to get back to the original $K$.
First adding the piece of stretched DNA, and then unwinding to
find the dependence on $L$ only, we find
\begin{align}
\F_{\coex}(K, L+\delta L) &= 
	\F_{\coex}(K,L) - \left( \frac{\tau^2}{2C} + \Feff  \right) \delta L  .
	\label{FLscaling}
\end{align}				
Similarly for the extension, [using $\xi(\tau)$ from Eq.~(1)]
\begin{align}
z_\coex(K,L+\delta L) &= 
		\label{zLscaling}
		z_\coex(K,L) + \left( \xi(\tau) + 
		 \frac{\tau}{2\pi C} q \right)\delta L.
\end{align}
Combining \eqrefTwo{FKscaling}{zKscaling} with
\eqrefTwo{FLscaling}{zLscaling}, we can write the free
energy and extension of the CS as linear in $K$ and
$L$, each with a slope and an intercept:
\begin{align}
\label{F_coex}
\F_{\coex}(K,L) &= \F_0 
	+ 2\pi \tau K - \left( \frac{\tau^2}{2C} + \Feff  \right) L ; \\
\label{z_coex}
z_{\coex}(K,L)  &= -z_0 
    - q K 
    + \left( \xi(\tau) + \frac{\tau}{2\pi C}q \right)L.
\end{align}
Note that $C$ and $\xi(\tau)$ are known from experiments on stretched DNA,
leaving the four anticipated force-dependent 
quantities to be described by a theory
of supercoiling: $\tau$, $q$, $\F_0$, and $z_0$.

\subsection{Self-repulsion}
It is essential to include a repulsive force between sections of the
DNA that come near each other; without it, the rod can pass
through itself, unphysically removing linking number in the process and
preventing the formation of plectonemes.  The physical origins of 
repulsive forces in DNA include both electrostatic and entropic effects.
We use discretized versions of the repulsive interactions 
described in Ref.~\cite{MarSig95a}.  

Electrostatic forces are modeled
using a Debye-Huckel screened Coulomb interaction:
\begin{equation}
E_{\mathrm{SC}}(r) = \frac{|e_- \nu d|^2}{\epsilon} \frac{e^{-r/\lambda_D}}{r},
\end{equation}
where $\nu = 8.4$ nm$^{-1}$ is the effective number of electron charges
per unit length, $\lambda_D = 0.8$ nm is the Debye screening length, and
$e_-^2/\epsilon = 2.9$ pN nm$^2$.  (These values are dependent on the
ionic concentration of the buffer, and were picked to match 
with $\approx 150$ mM NaCl.)

The entropic free energy of a helical structure is calculated in
Ref.~\cite{MarSig95a}, coming from the increasing confinement of 
fluctuations in more tightly coiled structures.  We use the same free
energy, written as a pairwise interaction between segments:
\begin{equation}
E_{\mathrm{ent}}(r) = \frac{2^{5/3} \sqrt{\pi} \Gamma(1/3)}{\Gamma(5/6)} 
		      \frac{kT d^2}{ (B/kT)^{1/3} r^{5/3} }.
\end{equation}

Since we also include straight parts of the DNA that should 
not have the same entropic
interaction, we cut off the entropic potential at a distance of 
$2 B / kT$, where the argument for the form of the potential 
breaks down \cite{MarSig95a}.


\subsection{Extra terms in the circular end-loop model}

Extra terms in the free energy that we have not considered would change
the predictions of the circular end-loop model --- these could include
electrostatic interactions, entropic effects, etc.  In fact, we can solve
for the properties that such an extra free energy term (call it $\Fex$)
would need to have in order to make the model match the experimental data.

\begin{figure}
\centering
\includegraphics[width=\little]{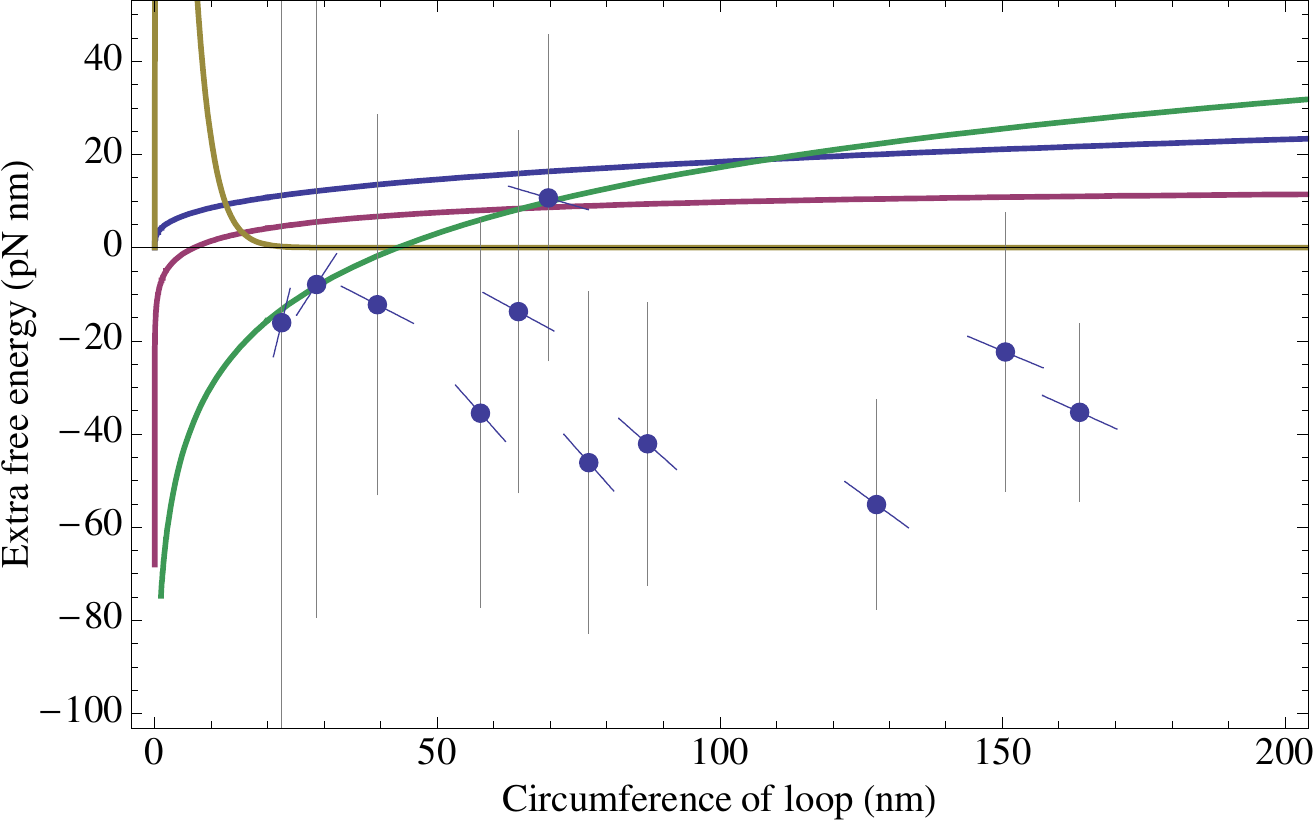}
\end{figure}

\begin{figure}
\centering
\includegraphics[width=\little]{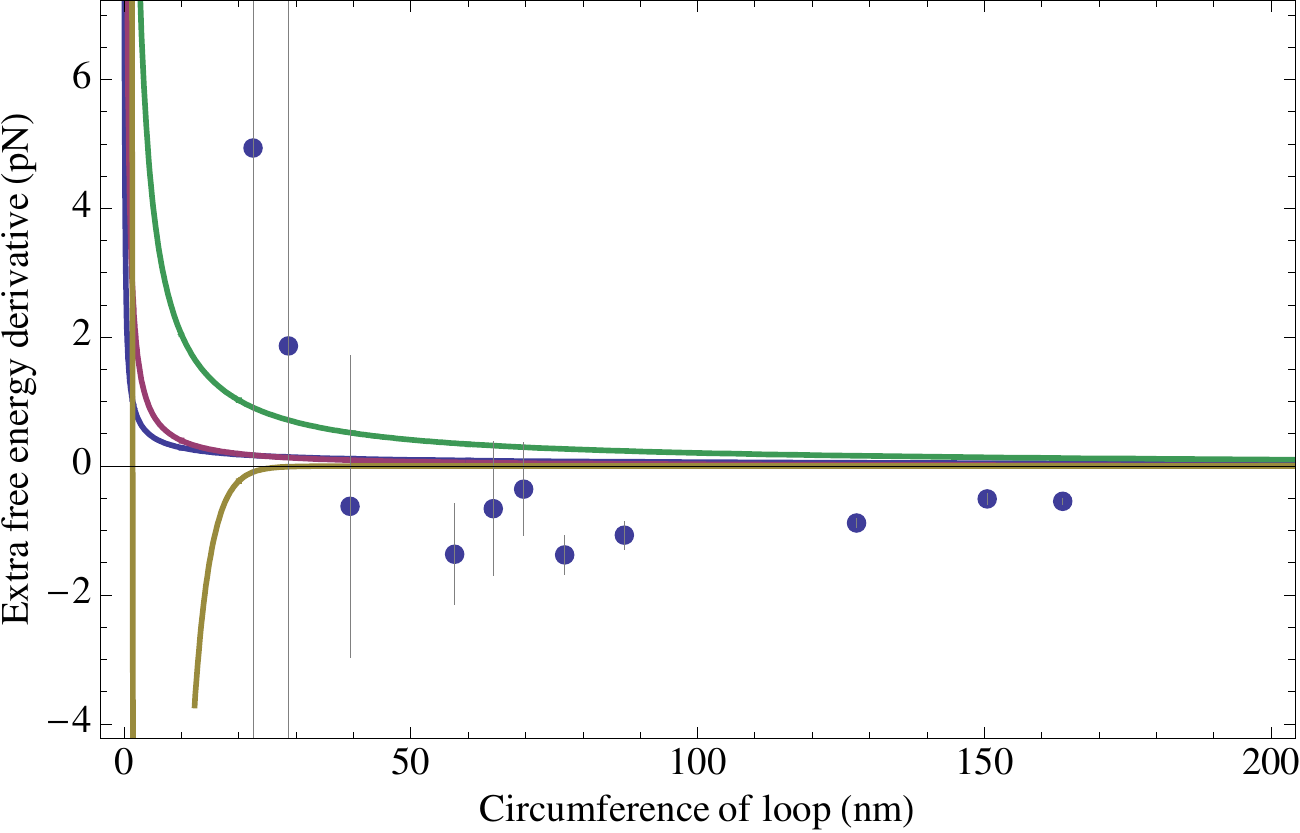}
\caption{\label{fig:EntropicTerms}
Entropic corrections from the literature do not help the circular
end-loop model fit the data.  The dots show the required free energy
contribution $\Fex$ (top plot) and its derivative with respect to end-loop
circumference $d\Fex/dL_l$ (bottom plot) that would produce an $\F_0$ and 
$z_0$ that match with the experiment (with $\Wr_\Loop=0.8$).  Bars on the top
plot show the required derivative, the value of which is shown on the bottom
plot.  Vertical grey lines show one standard deviation error bars.  Note
especially the inability of any of the proposed entropic terms to match the
well-constrained negative derivative at large end-loop circumferences 
(which happen at low force in the experiment); this
produces $L_l$ (and thus $z_0$) that are too small at low forces.  
A lessening of the effective force felt by the end-loop of about 0.5 pN
would help agreement, but none of the proposed corrections provides this.
}
\end{figure}

Adding this unknown term, we have
\begin{equation}
\F_l(K_l,L_l) = \frac{C}{2 L_l}\left[ 2\pi (K_l - \Wr_\Loop) \right]^2
	+ (2\pi)^2 \frac{B}{2 L_l} + \Fex(K_l,L_l).
\end{equation}
Since the terms we will imagine adding will not depend on
$K_l$, we will assume that $\Fex$ is only a function of $L_l$.
Setting the force and torque equal to the coexisting state values
($d\F_l/dL_l = -(\Feff +\tau^2/(2C))$; $d\F_l/dK_l = 2\pi\tau$)
then gives
\begin{align}
L_l^* & = 2\pi \sqrt{ \frac{B}{2(\Feff +d\Fex/dL_l)} } \\
K_l^* & = \frac{\tau L_l^*}{2\pi C} + \Wr_\Loop.
\end{align}
We now use the fact that
\begin{align}
\F_0 & = \F_l^* + (\Feff +\frac{\tau^2}{2C})L_l^* - 2\pi \tau K_l^* \\
z_0 & = \xi(\tau)L_l^* - q \left( K_l^* - \frac{\tau L_l^*}{2\pi C} \right)
\end{align}
to solve for the necessary values of $\Fex$ and $d\Fex/dL_l$ in order
to match with the experimental $\F_0$ and $z_0$.  We find
\begin{align}
\Fex & = \F_0 + 2\pi\tau \Wr_\Loop - \Feff  L_l^* - \frac{2\pi^2 B}{L_l^*}  \\
\frac{d\Fex}{dL_l} & = \frac{2\pi^2 B}{L_l^{*2}} - \Feff ,
\end{align}
where
\begin{equation}
L_l^* = \frac{z_0 + q \Wr_\Loop}{\xi(\tau)}.
\end{equation}
These required properties of the added free energy term are
plotted in \figref{fig:EntropicTerms} for $\Wr_\Loop = 0.8$.

We can then test whether different possible extra free energy terms
would match the requirements.  Here we try four possibilities taken
from the literature.
First, there is electrostatic repulsion coming from like charges
on opposite sides of the DNA circle.  This looks like (using the 
Debye-Huckel formulation from Ref.~\cite{MarSig95a})
\begin{equation}
\Fex^{\mathrm{electrostatic}} = 
	kT l_B \nu^2 K_0\left(\frac{L_l}{\pi \lambda_D}\right) L_l
\end{equation}
and is plotted in yellow in \figref{fig:EntropicTerms}.
Second, Odijk calculates the free energy for a circular DNA loop
and finds terms in the free energy \cite{Odi96} [Eq.~(2.13)]
\begin{equation}
\Fex^{\mathrm{Odijk}} = kT \log{ \frac{2\pi L}{B/(kT)} } - \frac{(kT)^2}{8 B} L;
\end{equation}
this is plotted in purple in \figref{fig:EntropicTerms}.
Third, a similar term is found by Tkachenko in solving for the J-factor
for unconstrained DNA cyclization 
\footnote{A. V. Tkachenko, q-bio/0703026 (2007).}
 [Eq.~(4)]:
\begin{equation}
\Fex^{\mathrm{Tkachenko}} = 5 kT \log \frac{L}{B/(kT)};
\end{equation}
this is plotted in green in \figref{fig:EntropicTerms}.
Finally, we could imagine that entropic contributions from confinement
similar to the one used by us for our elastic simulation could be important.
Although the form was derived for a different configuration (superhelical DNA),
we could try it to see if something similar might help.  Integrating the
confinement entropy from Marko and Siggia \cite{MarSig95a} over a circle gives
\begin{equation}
\Fex^{\mathrm{confinement}} = \frac{kT}{(B/kT)^{1/3} (L/(2\pi))^{2/3}} L,
\end{equation}
which is plotted in blue in \figref{fig:EntropicTerms}.

Although these possible terms are only initial guesses at the possible
corrections due to entropic and other effects, we see that they are all
qualitatively unable to help, especially at long loop lengths, which is
where the circular loop model fares worst at fitting the data.

\subsection{Calculating entropic contributions from fluctuations in 
plectoneme location, length, and linking number}
\label{appendix_entropy}

To investigate entropic effects, we would like to find the free energy 
of states with multiple plectonemes
\footnote{
	If the free energy necessary to nucleate a plectoneme is large
	compared to $kT$, then the coexisting state will contain 
	a single plectoneme.  If this is not the case, however (for example, when
	$L$ becomes large), we will need to consider equilibrium
	states in which multiple plectonemes coexist.
}, 
including fluctuations of linking number
and length both within individual plectonemes and moving among different
plectonemes.
We can achieve this by calculating the 
partition function for a state with $n$ plectonemes, identifying unique
states by the plectoneme positions $s_i$, the plectoneme lengths $L_{pi}$, and
the plectoneme linking numbers $K_{pi}$:
\begin{align}
Z_n(K,L) = & 
	\frac{1}{L_0^n} \int_0^L ds_1 \int_{s_1}^L ds_2 ... \int_{s_{n-1}}^L ds_n \\
	& \nonumber
	\frac{1}{L_0^n} \int_0^L dL_{p1} \int_0^L dL_{p2} ... \int_0^L dL_{pn} \\
	& \nonumber
	\frac{1}{K_0^n} \int_{-\infty}^{\infty} dK_{p1} 
			\int_{-\infty}^{\infty} dK_{p2} ... \int_{-\infty}^{\infty} dL_{Kn} \\
	& \nonumber
	\exp{ \left[ -\F_n(L,K,L_{pi},K_{pi})/kT \right] },
\end{align}
where we have neglected the complications coming from the possibility
that plectonemes could overlap.  The constants $L_0$ and $K_0$ set the length 
change and linking number change, respectively, 
that produce an independent state.  Since we are only concerned with the
free energy difference between the straight state and coexisting state, 
these constants would be set by the change in entropy of the degrees of
freedom in the straight state that are lost to the collective modes we
are integrating over in the coexisting state.

The first line of integrals represents the choice of where to put each
plectoneme, which does not change the free energy ($\F_n$ does not depend
on $s_i$).  We therefore simply get a factor of $L^n$, divided by $n!$
since plectonemes are indistinguishable:

\begin{equation}
Z_n(K,L) = \frac{(L/L_0)^n}{n!} \frac{1}{L_0^n K_0^n}
	\int_0^L \prod_i dL_{pi} 
	\int_{-\infty}^{\infty} \prod_i dK_{pi}  
	\exp{ \big[ -\F_n(L,K,\{L_{pi}\},\{K_{pi}\})/kT \big] }.
\end{equation}

Next we need to know the free energy of coexisting states that are away from
the equilibrium plectoneme length and linking number.  Assuming that the
plectoneme free energy density is quadratic in linking number density 
(as in Marko's model \cite{Mar07}), this turns out to be
\begin{align}
\F_n(L,K,\{L_{pi}\},\{K_{pi}\}) = & 
	\sum_{i=1}^n \frac{C}{2} \left( \frac{1}{1+v} \right)  
	\left( 2\pi \frac{K_{pi}}{L_{pi}} \right)^2 L_{pi} \\
	& \nonumber
	+ \frac{C}{2} \left( 2\pi \frac{K-\sum K_{pi}}{L-\sum L_{pi}} \right)^2
	(L - \sum L_{pi}) - \Feff (L - \sum L_{pi}) + n\mu ,
\end{align}
where $\mu$ is the chemical potential for plectoneme 
ends 
and $v \equiv 2 C \Feff  / \tau^2 $.

We first evaluate the integrals over $K_{pi}$, which amount to $n$ Gaussian
integrals; this gives 
\begin{align}
Z_n(K,L) = \frac{(L/L_0)^n}{n!} \frac{1}{L_0^n K_0^n} \pi^{n/2} 
	\int_0^L & \prod_i dL_{pi} 
	\left( \frac{ \prod_i L_{pi}/c_1 }
				{ 1 + (1+v)\frac{\sum L_{pi}}{L-\sum L_{pi}} } \right)^{1/2} \\
	& \nonumber
	\exp{ ( -\frac{1}{kT} \left[ 
		\frac{ \frac{C}{2} (2\pi K)^2 }{ L-\sum L_{pi} + (1+v)(\sum L_{pi})}
		-\Feff (L-\sum L_{pi}) + n\mu \right] ) }.
\end{align}
Now changing to unitless variables $x_i = L_{pi}/L_p$ and
$y = L_p/L$, and rearranging to 
move all the factors that depend on the sum of the plectoneme lengths
$y$ into the exponent, the term in the exponent becomes
\begin{equation}
f(y) =  \frac{1}{kT} \left( \frac{ \frac{C}{2} (2\pi K)^2/L }{ 1+vy }
		-\Feff L(1-y) + n\mu \right) 
		+\frac{1}{2} 
			\log{ \left( \frac{1+vy}{1-y} \right) },
\end{equation}
and we have
\begin{align}
Z_n(K,L) & = \frac{(L/L_0)^n}{n!} \frac{1}{L_0^n K_0^n} \pi^{n/2}  
	\int_0^L \prod_i dx_i 
	\sqrt{ \prod_i L_{pi}/c_1 }
	\exp{ [ - f(\sum L_{pi}/L) ] } \\
	& \nonumber
	= \frac{(L/L_0)^n}{n!} \frac{1}{L_0^n K_0^n} \pi^{n/2} 
	\int_0^L dL_p ~
	\delta \left( \sum L_{pi} - L_p \right)
	\int_0^{L_p} \prod_i dL_{pi}
	\sqrt{ \prod_i L_{pi}/c_1 }
	\exp{ [ - f(L_p/L) ] } \\
	& \nonumber
	= \frac{(L/L_0)^n}{n!} \frac{1}{L_0^n K_0^n} \pi^{n/2} 
	\int_0^L dL_p
	\frac{L_p^n}{L_p} \left( \frac{L_p}{c_1} \right)^{n/2}
	\left[ \int_0^1 \prod_i dx_i \sqrt{ \prod_i x_i } ~
		\delta\left( \sum x_i - 1 \right) \right]
	\exp{ [ - f(L_p/L) ] } \\
	& \nonumber
	= \frac{ (L/L_0)^{2n} (L/c_1)^{n/2}}{K_0^n} 
	  ~ \frac{ \pi^{n/2} ~ \gamma_n }{n!}
	\int_0^1 dy ~
	\exp{ [ - (f(y) - \frac{3n-2}{2} \log y) ] }.
\end{align}
The integral in large square brackets 
(characterizing fluctuations
in the individual plectoneme lengths that do not change the total
plectoneme length) gives a numerical 
constant $ \gamma_n = \pi^{n/2} / (2^n \Gamma(3n/2)) 
= 2^{\lfloor \frac{n-1}{2} \rfloor} \pi^{\lfloor \frac{n}{2} \rfloor}
/ (3n-2)!!$.  To evaluate the $y$ integral over total plectoneme 
length, we make a Gaussian approximation [noting that the total length 
is well-constrained by $f(y)$].  Then the fluctuations in the
(fractional) total length of plectonemic DNA are of size
\begin{equation}
\label{sigma_y}
\sigma_y = \left( 
	\frac{d^2}{dy^2}\left[f(y)-\frac{3n-2}{2}\log{y}\right] \Bigg |_{y^*}
			\right)^{-1/2},
\end{equation}
where $y^*$ is the equilibrium value of $y$, and the derivative is
\begin{equation}
\frac{d^2}{dy^2}\left[f(y)-\frac{3n-2}{2}\log{y}\right] = 
	\frac{1}{2}\left( \frac{1}{(1-y)^2} + \frac{3n-2}{y^2}
	-\frac{v^2 \left( 1 - \frac{8 \pi^2 C K^2}{L kT (1+vy)} \right)}
														{(1+vy)^2} \right).
\end{equation}
Without the entropic corrections, the
equilibrium length is $y^* = (u-1)/v$, where $u=2\pi C K/(\tau L)$.
We can safely use this value if we
are far from $y^*=0$ and $y^*=1$, and get
\begin{equation}
\label{sigma_y_approx}
\sigma_y = \frac{\sqrt{2}}{v}\left( \frac{1}{u}\frac{2 \tau^2 L}{kT C}
								- \frac{1}{u^2}
								+ \frac{1}{(v-u+1)^2} 
								+ \frac{3n-2}{(u-1)^2} \right) ^ {-1/2}.
\end{equation}
[Since we are usually near $y^*=0$ at the transition, to calculate
the length-dependence shown in 
Fig.~4 (left), we approximate $y^*$ numerically and use \eqref{sigma_y} 
instead of \eqref{sigma_y_approx}.]
In the end, we have
\begin{equation}
\label{Z}
Z_n(K,L) = \frac{ (L/L_0)^{2n} (L/c_1)^{n/2}}{K_0^n} 
	~ \frac{ \pi^{n/2} ~ \gamma_n }{n!}
	\sqrt{2\pi} \sigma_y 
	\left(\frac{u-1}{v}\right)^{(3n-2)/2}
	\left(\frac{v-u+1}{uv}\right)^{1/2}
	\exp{[-\F(K,L)/kT]}.
\end{equation}
The full partition function for all plectonemic states is then
\begin{equation}
Z(K,L) = \sum_{n=1}^{\infty} Z_n(K,L)
\end{equation}
(which we can numerically approximate by truncating the series 
at a reasonable $n$), such that the
coexisting state free energy is given by $\F_\coex(K,L) = -kT \log Z(K,L)$.
For the experimental values, we find that only the single plectoneme $n=1$
state contributes significantly near the transition.


\subsection{Independence of results on entropic effects}


In the paper, we have set the entropy from the previous section to zero
($S=0$) for most of the calculations.  How would we expect that including
$S$ would change any of the results?

First, $S$ would create a shift between the experimental $\F_0$ and the
predictions from models that do not include fluctuations.  We find that
this shift is largely 
independent of force, and is mostly dependent on $L_0$.  We do not
currently have a way of calculating $L_0$, but we expect that it should
be on the order of the persistence length of DNA, about 50 nm.
We find that setting $L_0$ to about 100 nm makes the prefactor equal
to 1, or equivalently sets $S=0$.  If we assume that $L_0$ is about
equal to the persistence length of DNA, we expect that we would need
to shift the model predictions by at most about 
$kT \log 2 \approx 5$ pN nm.

Second, we find that $S$ has a logarithmic dependence on $L$.  This means
that we expect $\F_0$ to decrease by something on the order of 
$ kT \log( L_2/L_1 ) $ when we increase the length from $L_1$ to $L_2$.
For the experimental lengths (with $L_2 \approx 2 L_1$), this again 
corresponds to a shift of about 5 pN nm.

Shifting $\F_0$ by these amounts would slightly change only 
the theory curves for $\F_0$ (about 5 pN nm), $\Delta z$ (about 10 nm), 
and $\Delta \tau$ (about 1 pN nm). 

\end{widetext}

\bibliographystyle{apsrev}
\bibliography{PlectonemePaper}